\documentclass[twocolumn,showpacs,preprintnumbers,amsmath,amssymb,prl,superscriptaddress]{revtex4}

\usepackage{graphicx}
\usepackage{dcolumn}
\usepackage{bm}

\begin{document}

\preprint{}

\title{Broadly Tunable Sub-terahertz Emission from Internal Branches of the Current-voltage Characteristics of Superconducting Bi$_2$Sr$_2$CaCu$_2$O$_{8+\delta }$ Single Crystals} 

\author{Manabu Tsujimoto}%
 \altaffiliation[Research Fellow of ]{the Japan Society for the Promotion of Science, 8 Ichiban-cho, Chiyoda-ku, Tokyo 102-8472, Japan}
\author{Takashi Yamamoto}%
 \altaffiliation[Present address: ]{Quantum Beam Science Directorate, Japan Atomic Energy Agency}
\author{Kaveh Delfanazari}%
\author{Ryo Nakayama}%
\author{Takeo Kitamura}%
\author{Masashi Sawamura}%
\author{Takanari Kashiwagi}
\author{Hidetoshi Minami}%
\author{Masashi Tachiki}%
\author{Kazuo Kadowaki}%
 \email{kadowaki@ims.tsukuba.ac.jp}

\affiliation{%
Graduate School of Pure \& Applied Sciences, University of Tsukuba,
1-1-1, Tennodai, Tsukuba, Ibaraki 305-8573, Japan
}%

\author{Richard A. Klemm}

\affiliation{Department of Physics, University of Central Florida, Orlando, Florida 32816, USA}%

\date{\today}

\begin{abstract}
Continuous, coherent sub-terahertz radiation arises when a dc voltage is applied across a stack of the many intrinsic Josephson junctions in a Bi$_2$Sr$_2$CaCu$_2$O$_{8+\delta }$ single crystal.  The active junctions produce an equal number of $I$-$V$ characteristic branches.  Each branch radiates at a slightly tunable frequency obeying the ac Josephson relation.  The overall output is broadly tunable and nearly independent of heating effects and internal cavity frequencies.  Amplification by a surrounding external cavity to allow for the development of a useful high-power source is proposed.
\end{abstract}

\pacs{07.57.Hm, 74.50.+r, 85.25.Cp}
\maketitle


Continuous, broadly tunable, and coherent sources of electromagnetic (EM) radiation are presently unavailable for frequencies within the ``terahertz gap'', the 0.3--10~THz range of crucial importance for many applications~\cite{Borak05,Tonouchi07}.  This gap can be filled by the ac Josephson effect intrinsic to a atomic-scale layered superconductor.  Application of a dc voltage $V$ across a single Josephson junction between identical superconductors leads to an ac Josephson current and EM radiation with the same frequency $f$ satisfying the ac Josephson relation, $f = f_{J} = \left( 2 e / h \right) V$, where $e$ is the electric charge and $h$ is Planck's constant~\cite{Josephson62,Langenberg65,Yanson65}.  Using two-dimensional arrays of Josephson junctions between wires of superconducting Nb, coherent radiation was observed when the array was placed parallel to a Nb ground plane~\cite{Barbara99}, that amplified the radiation, but did not affect its frequency, which obeyed the Josephson relation.  However, the technical problems involved in mass production were formidable.

The layered, high transition temperature $T_c$ superconductor, Bi$_2$Sr$_2$CaCu$_2$O$_{8+\delta }$ (Bi-2212), behaves as a stack of intrinsic Josephson junctions (IJJs)~\cite{Kleiner92}.  In Bi-2212, each of the junctions is naturally identical, as they are evenly spaced with two junctions per unit cell $c$-axis edge length of 1.533~nm.  Recently, continuous, coherent EM radiation was induced by applying a $V$ across the stack of $N$ IJJs present in small mesas milled out of single crystalline Bi-2212~\cite{Ozyuzer07,Kadowaki08,Minami09,Wang09,Ozyuzer09,Wang10,Guenon10,Kadowaki10,Tsujimoto10,Yamaki11,Kashiwagi11}.  The mesas typically have thicknesses $d \sim $ 1--2~$\mu $m and areas that vary from $\sim 4 \times 10^{-9}$~m$^2$ to $\sim 4 \times 10^{-8}$~m$^2$.  The thin mesa shapes were mostly rectangular, but some were square or circular~\cite{Tsujimoto10}.  Since Bi-2212 is extremely anisotropic, behaving for ${\bm E} \parallel \hat{\bm c}$ as an insulator~\cite{Kleiner92}, the three-dimensional mesa structure also behaves as an internal EM cavity, which couples to the non-linear ac Josephson currents generated in each junction~\cite{Ozyuzer07,Kadowaki08,Minami09,Wang09,Ozyuzer09,Wang10,Guenon10,Kadowaki10,Tsujimoto10,Yamaki11,Kashiwagi11}.  Besides satisfying the ac Josephson relation for a stack of $N$ IJJs, $f = f_{J} = \left( 2e/h \right) V / N$, it was also consistently reported that $f$ locked onto an internal cavity mode frequency $f^c_{m,p}$, where $m$ and $p$ are integers appropriate for the geometry.  Neglecting heating effects, for a very thin ($d \ll w \le \ell$) rectangular cavity of length $\ell$ and width $w$, the relevant transverse magnetic TM$^z$($m$,$p$) modes have frequencies $f^c_{m,p} = \left( c_0 / 2n \right) \sqrt{ \left( m/w \right) ^2 + \left( p/\ell \right) ^2}$, where $c_0$ is the speed of light in vacuum and $n=\sqrt{\epsilon } \approx 4.2$ is the index of refraction for ${\bm E} \parallel \hat{\bm c}$ in Bi-2212~\cite{Ozyuzer07,Kadowaki08,Minami09,Wang09,Ozyuzer09,Wang10,Guenon10,Kadowaki10,Tsujimoto10,Yamaki11,Kashiwagi11,Benseman11,Bulaevskii07,Lin08-2,Tachiki09,Klemm10,Klemm11}.

\begin{figure}[t] 
	\includegraphics[width=0.48\textwidth ,clip]{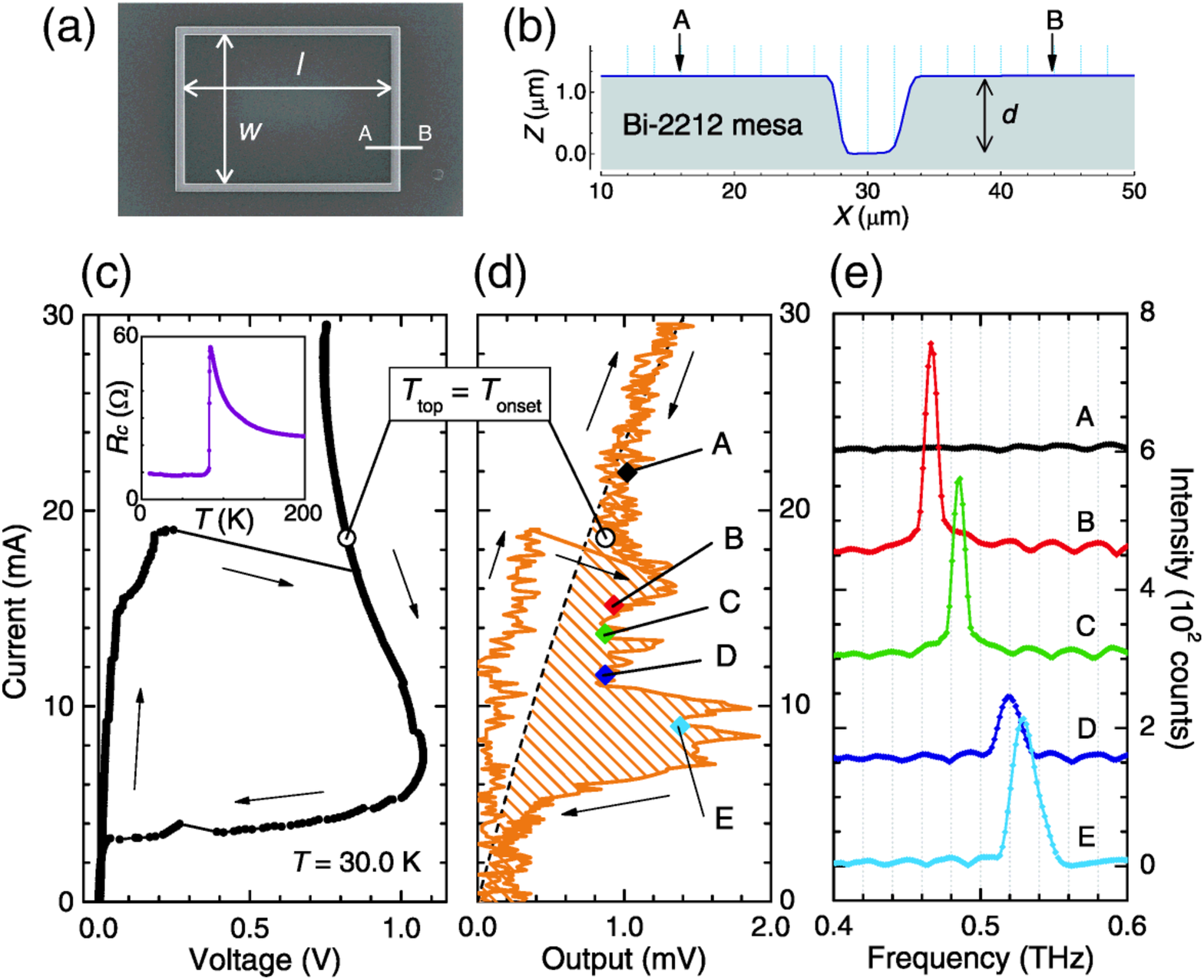} \\
	\caption{(color online)
R1 structure, $c$-axis resistivity, outermost-branch IVC, emission output, and frequency spectra.  (a) SIM image.  (b) AFM image along A--B in Fig.~1(a).  (c) Outer-branch IVC.  Inset: $R_c \left( T \right) $. (d) Plot of $I \left( V_{\rm out} \right)$ corresponding to Fig.~1(c).  Open circles: $T_{\rm{top}} = T_{\rm{onset}} = 84.4$~K.  (e) Spectral intensities from points A--E in Fig.~1(d).
}
\end{figure}%

Most workers have thought the enhancement of the output radiation by the excitation of an internal cavity mode was so strong that the radiation from the ac Josephson current source alone was too weak to observe~\cite{Ozyuzer07,Kadowaki08,Minami09,Wang09,Ozyuzer09,Wang10,Guenon10,Bulaevskii07,Lin08-2,Tachiki09}.  However, recently the contributions to the output power from the ac Josephson current source alone and that enhanced by resonance with an internal EM cavity source were found to be comparable in magnitude~\cite{Kadowaki10,Tsujimoto10,Klemm10,Klemm11}.  This created a great deal of controversy.  Here we show clear evidence that the mesas can emit radiation at many frequencies, without strong interaction with an internal EM cavity mode.  More important, the resulting radiation is tunable over a broad range of frequencies, allowing for the construction of a powerful device that could fill the terahertz gap.

\begin{figure}[t] 
	\includegraphics[width=0.48\textwidth ,clip]{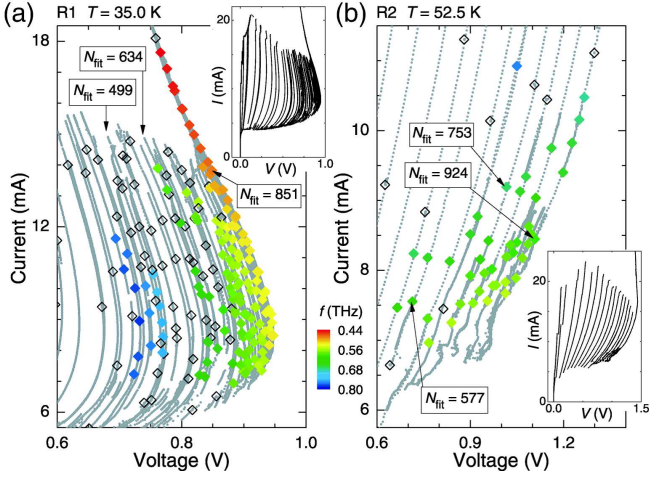} \\
	\caption{(color online)
Emission from IVC points at the color-coded frequencies was observed at the filled diamonds.  No radiation was detected at the open diamonds.  Arrows indicate the numbers $N_{\rm fit}$ of resistive junctions from fits to the ac Josephson relation.  (a) R1. (b) R2.  Insets: Full IVCs.
}
\end{figure}%

We studied two rectangular mesas, R1 and R2.  R1 was prepared by milling a groove into the Bi-2212 substrate, but R2 was sandwiched between two Au layers.  Further information on R1, R2, and  three other mesas of various shapes are presented in the Supplemental Material~\cite{SM}.  Figure 1(a) is a scanning ion microscope (SIM) picture of the top of R1, with $w = 99.2$--$102~\mu$m, $\ell = 137$--$140~\mu$m, and groove depth $d \sim 1.3$~$\mu $m, before the electrode attachments.   R1 is $\sim$~3\% longer and wider at the bottom of the groove than at the top.  Figure 1(b) is an atomic force microscope (AFM) scan of the groove profile along the A--B line in Fig.~1(a).  From this $d$ value, we estimate the number of IJJs in the stack to be $N_{\rm{max}} \sim 850$.  The hysteresis loop indicated by the arrows in Fig.~1(c) is typical of the outermost branch of the $I$-$V$ characteristics (IVCs) of an IJJ system.  At the largest jump from $V = 0.25$ to 0.85~V, all of the $N_{\rm{max}}$ junctions switch into the resistive state simultaneously.  In the high-$V$ bias region, R1 is inevitably Joule heated at a rate of $\sim 16$~mW.  This huge power dissipation may cause the negative differential resistance between $I = 7$ and 25~mA, since the $c$-axis quasiparticle resistance $R_c$ has a strong $T$-dependence as shown in the inset of Fig.~1(c).  On the return region of the $I$-$V$ loop where $I \sim 5$~mA, we observed several small steps, indicating that some of the IJJs make the transition from the resistive to the superconducting state.

Figure 1(d) shows the R1 radiation output intensity $V_{\rm{out}}$ generated from $I$ on the same scale as in Fig.~1(c).  The dashed curve indicates the expected drift due to thermal radiation from the sample and its holder.  Soon after the largest jump in the $I$-$V$ loop, intense emission of EM waves was clearly observed.  Strong spatial variation of the temperature $T_{\rm{top}}$ over the top mesa surface was previously observed~\cite{Wang09,Wang10,Guenon10}, and was thought to strongly affect the cavity mode-locking condition.  We take $T_{\rm{top}}$ to be the spatial average of the temperature on the mesa top.  For R1, $T_c = 81.1$~K with a width $\Delta T_c = 6.6$~K.  Below $T_{\rm{top}} = T_{\rm{onset}} =84.4$~K, indicated by the open circles in Figs.~1(c) and 1(d), R1 makes the transition to the superconducting state, as shown in Fig.~1(d).

Figure 1(e) shows the radiation spectra, offset by 150 counts for clarity, measured at points A--E  in Fig.~1(d), at which all of the $N_{\rm{max}}$ IJJs are in the resistive state.  The narrow and intense peaks in the emission spectra of  B--E have maxima varying from 0.47~THz to 0.53~THz.  The detector resolution-limited widths of the B, C peaks are too narrow to arise from synchronization by an internal cavity mode alone, and might involve heating effects~\cite{Wang10}.  Such outermost branch $f$ tunability of up to 40\% was found previously  by varying both $V$ and the bath $T$~\cite{Wang10,Kashiwagi11,Benseman11}.  Although the wide $f$ range strongly violates the internal cavity resonance condition, $f = f^c_{m,p}$,   the ac Josephson relation for a stack of $N_{\rm max}$ resistive IJJs~\cite{Wang10}, $f = f_{J} = \left( 2e/h \right)  V/N_{\rm max}$,  is excellently obeyed.

\begin{figure}[t] 
	\includegraphics[width=0.48\textwidth ,clip]{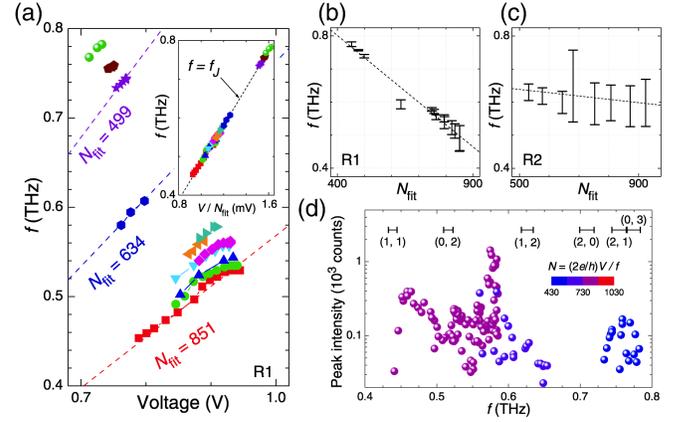} \\
	\caption{(color online)
Emission frequencies $f$ and intensities from R1, and $f$ versus resistive junction numbers $N_{\rm fit}$ for R1 and R2.  (a)  $f \left( V \right)$ plots for 11 R1 IVC branches. Dashed lines:  Fits to $f=f_J=(2e/h)V/N_{\rm fit}$.  Inset: $f(V/N_{\rm fit})$ plot for all R1 data. Dotted line: $f = f_{J}$.  (b), (c)  $f \left( N_{\rm{fit}} \right)$ ranges for R1 and R2, with  dotted  eye guides.  (d) Peak intensity on a logarithmic scale versus $f$ for R1.  Calculated $f^c_{m,p}$ ranges are given at the top.  $N_{\rm{fit}}$ values are color coded.}
\end{figure}%

We found that EM emission also occurs at many points in the inner region of the multiply-branched IVCs, where $N = 1, 2, \ldots, N_{\rm{max}}$ is fixed but different for each branch.  The synchronization of the $N$ IJJs on a single $I$-$V$ branch is also not controlled by an internal cavity resonance~\cite{Balanov09}, unlike some predictions~\cite{Koshelev10}.  In Figs.~2(a) and 2(b), the radiation frequencies, $f$, are plotted as color-scaled symbols on the high-bias regions of the multiply-branched $I$-$V$ structures for R1 at 35.0~K and R2 at 52.5~K, respectively.  The insets show the full IVCs.  All of the R1 curves in Fig.~2(a) bend backwards with increasing current, indicative of Joule heating~\cite{Wang09,Wang10,Guenon10,Kurter10}.  However, R2 is less susceptible to heating effects, and its IVCs in Fig.~2(b) are monotonic.  The $f$ spectra were obtained at as many $\left( I,V \right)$ bias points as possible.  At the bias points denoted by open diamonds, no emission was detected.

By repeated measurements of the emission from a particular $I$-$V$ branch with constant $N$, we confirmed that $f$ satisfies the ac Josephson relation $f = f_{J} = \left( 2 e / h \right) V / N$.  In Figs.~3(a) and 4(a), we replotted the emission data from the IVCs of R1 and R2 shown in Figs.~2(a) and~2(b) in terms of $f \left( V \right)$, representing the data from each branch in terms of unique symbols and colors.  By fits of the data for a particular branch to the ac Josephson relation with $N = N_{\rm fit}$, the experimental best-fit value $N_{\rm fit}$ for each emitting branch was determined.  Three examples each for R1 and R2 are respectively indicated by the dashed lines in Figs.~3(a) and 4(a) and the arrows in Figs.~2(a) and 2(b).  In the insets to Figs.~3(a) and 4(a), the entire emission data from the inner IVC branches  of R1 and R2 are respectively plotted as $f \left( V / N_{\rm fit} \right)$.  For both R1 and R2, the ac Josephson relation $f = f_{J} = \left( 2 e / h \right) V / N_{\rm fit}$, represented by each dotted line, is  well obeyed.

Since Figs.~3(a) and 4(a) clearly demonstrate that $f$ is slightly tunable on each branch (indicated by a fixed symbol and color), in Figs.~3(b) and 3(c), we respectively replotted the data from each R1 and R2 branch as $f \left( N_{\rm fit} \right)$.  Although for both samples $f$ is tunable as indicated by the vertical bars for each fixed branch number $N_{\rm fit}$, R1 and R2 display rather different aspects of tunability.  The overall tunability of R1 and R2, which are respectively tunable from 0.44~THz~$\le f \le$~0.78~THz and from 0.43~THz$\le f \le$0.76~THz, is nearly the same.  However, for R1, this range is primarily due to the $f$-dependence on $I$-$V$ branch number, whereas for R2, the tunability is greatest on a single branch.

We measured the spectrum for each emitting point in the inner IVC branches  of R1 and R2, and determined its peak intensity as  for points B--E on the outer branch of R1 shown in Fig.~1(e).  In Figs.~3(d) and 4(b), we plotted the peak intensity of the emissions on a logarithmic scale versus $f$ for R1 and R2, respectively, and $N_{\rm fit}$ is approximately coded with color.  The ranges due to mesa profile variations of the respective internal cavity mode frequencies $f^c_{m,p}$ are indicated at the figure tops.  For R1, no emission was found for $f < 0.40$~THz, excluding the expected cavity resonance frequencies of 0.255 and 0.357~THz corresponding to the TM$^z$(1,0) and TM$^z$(0,1) modes.  Moreover, the spectrum observed in Fig.~3(d) is almost completely unrelated to any of the internal cavity modes.  For example, the strongest intensity observed for $f \approx 0.575$~THz, corresponding to $V = 0.894$~V, $I = 9.56$~mA, and $N_{\rm fit} = 753$, is far from the two nearest cavity resonance frequencies.  Although the indicated cavity frequency ranges could be shifted to higher frequencies by shorter effective $\ell$ or $d$ values due to hot spots~\cite{Wang10}, since $T$ = 35~K in each measurement, such shifts could still not explain the broad range of observed frequencies.  For $N_{\rm max} = 851$, there are $\sim 4.3 \times 10^{130}$ ways to have $N_{\rm fit}$ of the $N_{\rm max}$ junctions in the resistive state, each of which could in principle lead to a different emission intensity.  Hence, multiply cycling through the IVCs led to a large variation in peak intensity at that $(I,V)$ point.  The synchronization of the $N_{\rm fit} < N_{\rm max}$ resistive junctions thus occurs independently of the internal cavity mode excitations~\cite{Balanov09}.  More important, the emission spectrum in Fig.~3(d) is continuous from 0.44~THz~$\le f \le$~0.78~THz, except for a gap between 0.66~THz and 0.73~THz.  The gap may be due to some experimental difficulty in accessing the appropriate branches.  This widely continuous spectrum independent of internal cavity resonances implies that $f$ is broadly tunable.  Figure 3(b) suggests that the $f$ values from lower branch-number emissions could exceed 1~THz.  Further tunability might arise from base $T$ variations~\cite{Wang10}, as discussed in the Supplemental Material~\cite{SM}.

The peak intensity spectrum of R2 measured at 52.5~K shown in Fig.~4(b) is also broadly tunable, with emission for 0.43~THz~${\le f \le}$~0.76~THz, and a small gap between 0.67 and 0.75~THz.  For R2, the $f$ range from 0.43~THz to $\sim$~0.55~THz is below the range of the lowest cavity resonance frequency $f^c_{1,0}$ values.  Unlike the peak intensity spectrum of R1, the largest intensity at 0.565~THz is close to this $f^c_{1,0}$ range.  In this case, the synchronization of the $N$ resistive junctions could be aided by the excitation of this internal cavity mode~\cite{Balanov09,Koshelev10}.

We measured the full angular dependence of the radiation from R2 on the outermost branch at $f = 0.639$~THz, and the results are shown and discussed in the Supplemental Material~\cite{SM}.  Although the observed angular dependence is similar to that expected from the low-$Q$ tail of the TM$^z$(1,0) cavity mode excitation, the frequency is far from $f^c_{1,0}$, but could be amplified by excitation of a higher frequency mode, as indicated.  We thus determined~\cite{Kadowaki10,Tsujimoto10,Klemm10,Klemm11} that the angularly integrated output power at this frequency is $\sim 0.2~\mu$W, somewhat smaller than previously reported~\cite{Ozyuzer07,Kadowaki08,Minami09,Wang09,Ozyuzer09,Wang10,Guenon10,Kadowaki10,Tsujimoto10,Yamaki11,Kashiwagi11,Benseman11}.  In addition, a much larger emission peak intensity, with an estimated overall emission power of 4.8~$\mu$W, was found at 0.557~THz, which is much closer to $f^c_{1,0}$.  Besides the excitation of the TM$^z$(1,0) mode, this strong inner branch emission could have been enhanced by R2 being sandwiched between two Au layers~\cite{Klemm10}.  More important, the large range in observed $f$ values well below the $f^c_{1,0}$ range is clear evidence that the highly tunable emission itself can occur without excitation of an internal cavity mode.

Although R1 was prepared by forming a groove into the Bi-2212 substrate,  R2 stood atop a Au substrate.  The heating effects and EM boundary conditions for R1 might be significantly different from those for R2~\cite{Koshelev10}.  Although the synchronization of the $N_{\rm fit}$ junction emissions from R2 is likely to be enhanced by internal cavity resonances~\cite{Bulaevskii07}, synchronization of the $N_{\rm fit}$ emissions from R1 could arise from the radiation itself~\cite{Balanov09}, but is more likely to be enhanced by the shunt capacitance in the electrical circuit arising from the non-emitting insulating junctions in the adjacent Bi-2212 substrate~\cite{Martin10}.

\begin{figure}[t] 
	\includegraphics[width=0.48\textwidth ,clip]{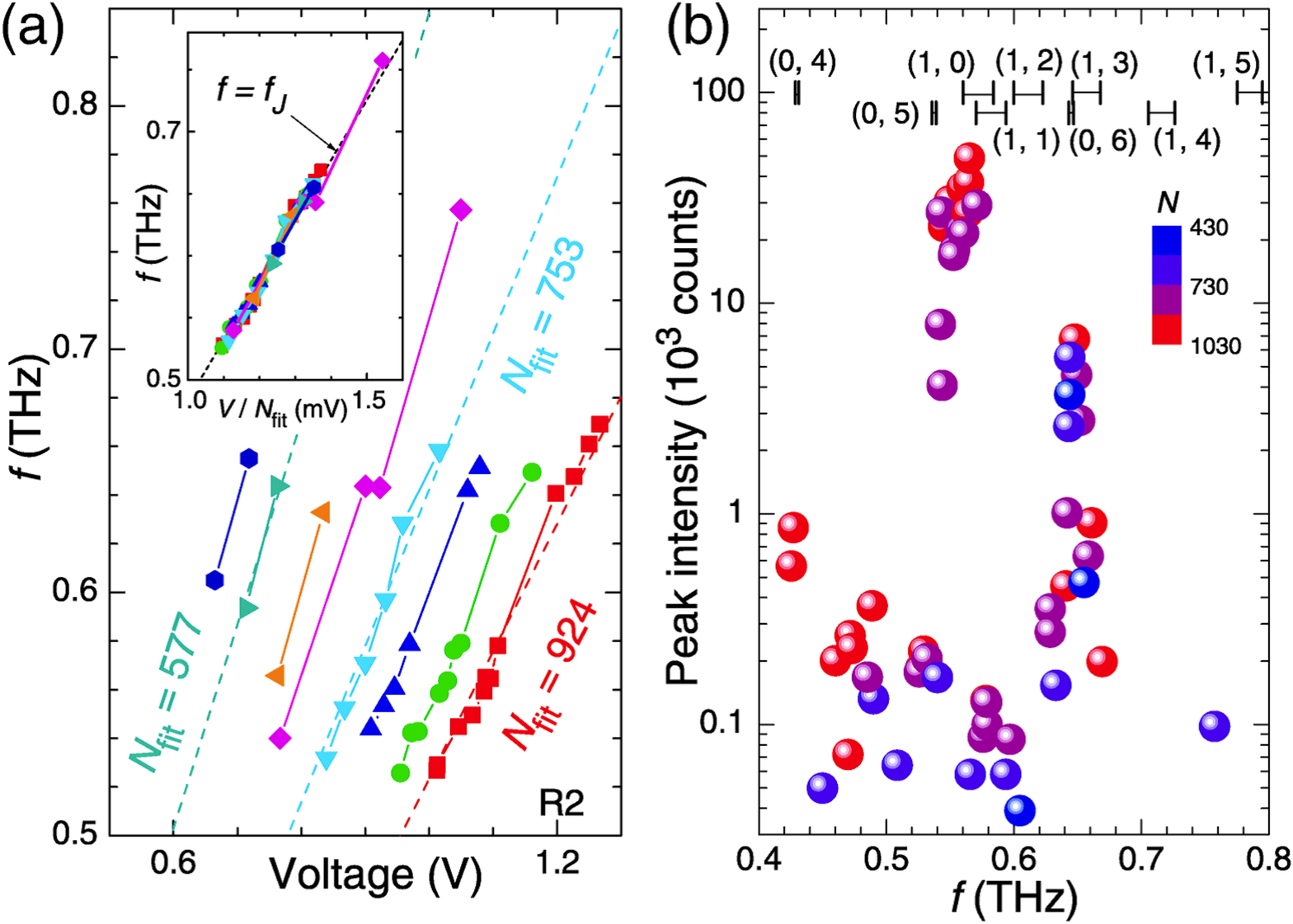} \\
	\caption{(color online)
Emission frequencies and intensities from R2.  (a) Plots of $f \left( V \right)$ for 8 branches in the R2 IVCs.  Dashed lines: Fits to $f = f_{J} = \left( 2e/h \right) V/N_{\rm{fit}}$.  Inset: Plot of $f \left( V/N_{\rm fit} \right)$ for all R2 data.  Dotted line: $f = f_{J}$.  (b) Peak intensity on a logarithmic scale versus $f$ for R2.  Calculated $f^c_{m,p}$ ranges are given at the top.  $N_{\rm fit}$ values are color coded.
}
\end{figure}%

To determine whether the excitation of an internal cavity mode was an essential feature of the coherent radiation obtained from mesas of Bi-2212 under the application of $V$ across the stack of IJJs in the mesa, we examined the inner branches of the IVCs.  We found these internal branches to emit radiation over a broad frequency range.  Hence, we conclude that the primary source of the intense, coherent sub-THz radiation is the ac Josephson current, and that the internal EM cavity produced by the geometrical shape of the emitting mesa is at best of minor importance, and for one sample, completely irrelevant.  Hence, broadband, tunable, continuous coherent radiation can be obtained in the sub-THz $f$ range from Bi-2212 mesas.  Since the wavelength in vacuum is $n \approx 4.2$ times longer than in the mesa, we propose that a high-power device can be constructed by surrounding the mesa with an external, tunable high-$Q$ EM cavity~\cite{Black01,Clark04}.  By examining yet lower inner branches, it ought to be possible to increase the upper $f$ limit into the 1--10 THz range.  Hence, it is now relatively straightforward to produce broadly tunable, continuous, coherent radiation over the  range of the terahertz gap.

The authors thank X. Hu, S. Lin, A. Koshelev, M. Matsumoto, T. Koyama, M. Machida, H. Asai, S. Fukuya, and K. Ivanovic for valuable discussions and Y. Ootuka, T. Hattori, A. Kanda, I. Kakeya, B. Markovic, K. Yamaki, and H. Yamaguchi for technical assistance.  This work was supported in part by CREST-JST (Japan Science and Technology Agency), WPI (World Premier International Research Center Initiative)-MANA (Materials Nanoarchitectonics) project (NIMS) and Strategic Initiative category (A) at the University of Tsukuba.

\end{document}